\documentclass[fleqn,twoside]{article}
\usepackage{espcrc2}
\usepackage{graphics}
\def\lsim{\raise0.3ex\hbox{$<$\kern-0.75em\raise-1.1ex\hbox{$\sim$}}}
\def\gsim{\raise0.3ex\hbox{$>$\kern-0.75em\raise-1.1ex\hbox{$\sim$}}}

\newcommand{\pslash}{p\kern-1ex /}
\newcommand{\Dslash}{{\cal D}\kern-1.5ex /}

\newcommand{\beqa}{\begin{eqnarray}}
\newcommand{\eeqa}{\end{eqnarray}}

\newcommand{\beq}{\begin{equation}}
\newcommand{\eeq}{\end{equation}}
\newcommand{\bc}{\begin{center}}
\newcommand{\ec}{\end{center}}

\title{
Two-particle wave function
in four dimensional Ising model
\author{
  T.~Yamazaki\address{RIKEN BNL Research Center, Brookhaven 
National Laboratory, Upton, NY 11973, U.S.A.}
}
}
\begin{document}
\pagestyle{empty}

\begin{abstract}
An exploratory study of two-particle wave function
is carried out with a four dimensional simple model.
The wave functions not only for two-particle ground and first excited 
states but also
for an unstable state are calculated from three- and four-point functions
using the diagonalization method suggested by L\"uscher and Wolff.
The scattering phase shift is
evaluated from these wave functions.
\end{abstract}

\maketitle

\section{Introduction}
\label{sec:intro}

Calculation of the scattering phase shift is
important to understand scatterings and decays of hadrons.
The phase shift in $I=2$ $\pi\pi$ scattering system~\cite{PSQCD}
was evaluated with the finite volume method~\cite{FVFP,FVFW} 
derived by L\"uscher.
In the finite volume method two-particle wave function plays 
an important role, because the phase shift is extracted by the wave function.

Study of the wave function 
was carried out by Balog {\it et al.}~\cite{Balog}.
Using two-dimensional statistical model
they evaluated the phase shift from the wave function for several 
two-particle states.
Recently Ishizuka {\it et al.} and CP-PACS collaboration
studied the wave function in $I=2$ $\pi\pi$ scattering system~\cite{WF}.
They extracted the scattering length and scattering effective potential
from the wave function for the ground state.
Here calculation of the wave function for 
the first excited and an unstable states is attempted
with a four dimensional simple Ising model.
It is also aimed to evaluate the phase shift from these wave functions.

\section{Wave function}

L\"uscher proved that the wave function $\phi( \vec{r} )$ 
satisfies effective Shr\"odinger equation,\\
$\displaystyle{
(\nabla^2 + p^2) \phi( \vec{r} ) = 
\int d^3 r^{'} U_E ( r, r^{'} ) \phi( r^{'} ),
}$\hfill(1)\\
where $\vec{ r }$ is relative coordinate of two particles,
$p$ is the relative momentum, and $U_E ( r, r^{'} )$ is
the Fourier transform of the modified Bethe-Salpeter kernel
introduced in ref.~\cite{FVFP}.
The effective potential $U_E ( r, r^{'} )$ depends on energy $E$
and decays exponentially in $r$ and $r^{'}$.
The wave function satisfies the Helmholtz equation
$(\nabla^2 + p^2) \phi( \vec{r} ) = 0$ in $r > R$.
The $R$ is the effective range where
$U_E ( r, r^{'} )$ becomes sufficiently small in exterior region of $R$.
L\"uscher found general solution of the Helmholtz equation in
a finite volume,\\ 
$\displaystyle{
G(\vec{r}) = (1/L^3)\sum_{\vec{k}}e^{i\vec{k}\cdot\vec{r}}/(\vec{k}^2-p^2)
}
$,\hfill(2)\\
where $\vec{k}=(2\pi/L)\cdot\vec{n}$ and $\vec{n}$ is an integer vector.
We can extract $p^2$ by fitting $\phi(\vec{r})$ in $r > R$ with $G(\vec{r})$,
and can then evaluate the phase shift with the finite volume method.

\section{Methods}

A simple model, which is constructed with a lighter mass field $ \pi $
coupled to a heavier mass field $ \sigma $
with three-point coupling,
is employed to calculate an unstable state.
This model has been successfully used to observe a resonance
by Gattringer and Lang~\cite{GatLan} 
in two dimensions, 
and by Rummukainen and Gottlieb~\cite{RG} in four dimensions.

Three- and four-point functions,
$F_i(\vec{r},t) = 
\langle 0 | \mathcal{ W } ( \vec{ r } ) 
[ \mathcal{ O }_i ( t ) - \mathcal{ O }_i ( t + 1 ) ] | 0
\rangle $ 
and $C_{ij}( t ) = 
\langle 0 | \mathcal{ O }_i ( 0 ) 
[ \mathcal{ O }_j ( t ) - \mathcal{ O }_j ( t + 1 ) ] | 0
\rangle $ for $ i,j = 0, 1 $ and $ \sigma $,
are calculated to obtain the wave function for some states.
The operator 
$\mathcal{ O }_i ( t )$ is $\pi\pi$ operator with $ p = 0, (2\pi/L)$ 
($ i = 0, 1$) and $\sigma$ operator ($i = \sigma$),
and the subtraction $[ \mathcal{ O }_i ( t ) - \mathcal{ O }_i ( t + 1 ) ]$
is performed to eliminate the vacuum contribution.
The wave function operator $\mathcal{ W } ( \vec{ r } )$ is defined by
$\mathcal{ W } ( \vec{ r } ) = 
(1/L^3)\sum_{\vec{X}, {\bf R}}\pi({\bf R}[\vec{r}]+\vec{X})\pi(\vec{X})$,
and the ${\bf R}$ is an element of cubic group, and summation over 
$\vec{X}$ and ${\bf R}$ projects to A$^+$ sector.
The lattice size is $L^3\times T = 20^3 \times 64$
and the number of the configurations is 405$\times 10^3$.
In this work the energy of the first excited state is larger than
that of the $\sigma$ state.

The wave function is defined by
$
\phi_\alpha( \vec{ r } ) 
= \langle 0 | \mathcal{ W }( \vec{ r } ) | \alpha \rangle
$ for the $\alpha = 0, 1$ and $\sigma$ state.
Using $ \phi_\alpha( \vec{ r } ) $
and the state overlap of the operator 
$V_{\alpha i } = \langle \alpha | \mathcal{ O }_i | 0 \rangle$,
one has 
$F_i(\vec{r},t) = \sum_\alpha 
\phi_\alpha(\vec{r})\Delta_\alpha(t) V_{\alpha i}$,
where $\Delta_\alpha ( t ) = (1 - e^{-E_\alpha})\,e^{- E_\alpha t} 
+ (1-e^{E_\alpha})\,e^{-E_\alpha (T-t)}$.
It is also easy to show that
$C_{ij}(t) = \sum_{\alpha} V^t_{i\alpha}\Delta_\alpha( t ) V_{\alpha j}$.
The assumption of this calculation is that 
higher energy states than the first excited state
do not contribute to $C_{ij}(t)$ and $F_i(\vec{r},t)$.
To obtain the wave function,
at first the overlaps are extracted 
by the diagonalization method~\cite{diag} with the correlation function matrix 
$C^{-1/2}(t_0) C(t) C^{-1/2}(t_0)$, where $t_0$ is reference point.
The diagonalization also yields the energy $E_\alpha$.
Then the wave function is extracted by the projection\\
$\displaystyle{
\phi_\alpha ( \vec{ r } ) = 
\Bigl[\sum_i F_{i} ( \vec{r},t ) V^{-1}_{i\alpha}\Bigr] / 
\Bigl|\sum_i F_{i} ( \vec{r}_0,t ) V^{-1}_{i\alpha}\Bigr|
}$\hfill(3)\\
apart from the normalization, where 
the normalization point $r_0$ is chosen $r_0 = 1$.

\begin{figure}[t!]
\centerline{\scalebox{.34}[.39]{\includegraphics{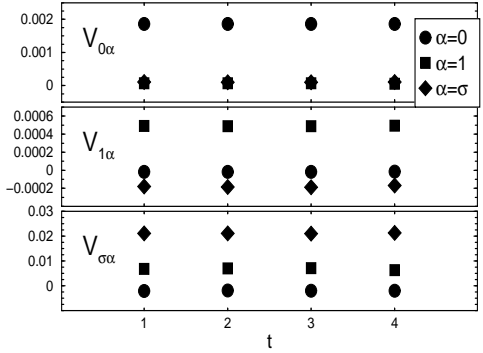}}}
\vspace{-10mm}
\caption{\label{fig:olap}
State overlaps of operators at several $t$.
}
\vspace{-6mm}
\end{figure}

\begin{figure}[t!]
\centerline{\scalebox{.33}[.33]{\includegraphics{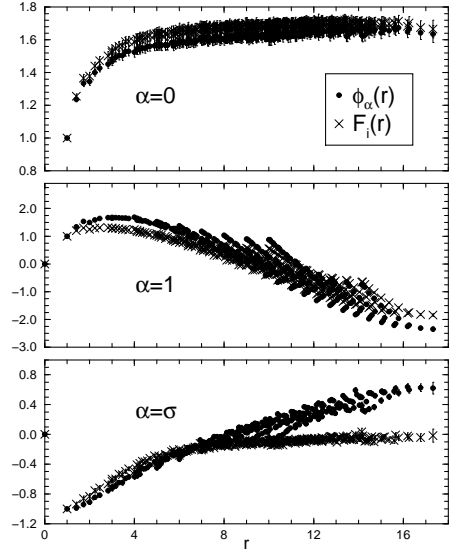}}}
\vspace{-10mm}
\caption{\label{fig:WF}
Wave function calculated from eq.~(3) for ground, first excited and
$\sigma$ states. $F_i(r)$ corresponding to each state is also presented.
}
\vspace{-6mm}
\end{figure}

\section{Results}

Fig.~\ref{fig:olap} displays the overlaps determined at some $t$.
The $ i = 0 $ operator is almost dominated by the ground state,
while other operators has contributions from each state.
All the overlaps are stable at small $t$ region,
so that $V_{i\alpha}$ is determined at $t = 1 $.
Using the overlaps 
it is possible to calculate the wave function not only for the 
first excited state but also for the $\sigma$ state 
as shown in Fig.~\ref{fig:WF}.
The projection eq.~(3) is carried out with $F_i( \vec{r}, t )$ at
$t = 3$ for $\alpha = 0,1$, and at $t=4$ for $\alpha = \sigma$.
The function $F_i( r, t )$ normalized at $r_0$ is also displayed in the figure
for comparing ones before and after the projection.
The difference between $\phi_\alpha( r )$ and $F_i( r, t )$
for the ground state is small, 
while those for the first excited and $\sigma$ states are large, 
as expected from the operator overlaps in Fig.~\ref{fig:olap}.

\begin{figure}[t!]
\centerline{\scalebox{.33}[.33]{\includegraphics{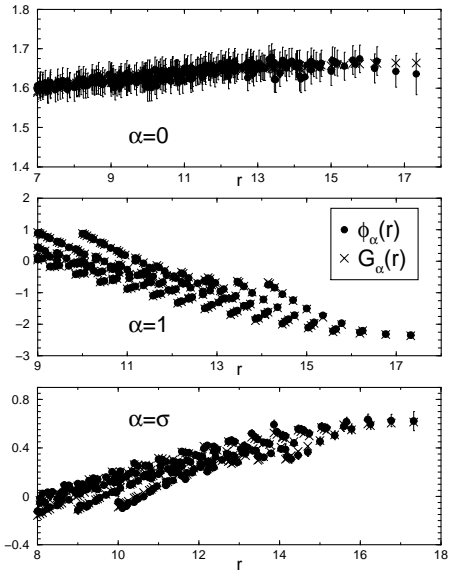}}}
\vspace{-10mm}
\caption{\label{fig:fit}
Fitting result of wave functions for ground, first excited and
$\sigma$ states.
}
\vspace{-6mm}
\end{figure}

The effective range is required for fitting of $\phi_\alpha( r )$
in $r > R$ using eq.~(2).
In Ref.~\cite{WF} the effective range is estimated from the quantity 
$( \nabla^2 \phi_\alpha( \vec{r} ) ) / \phi_\alpha( \vec{ r } )$,
which is approximately effective potential.
However, the effective range cannot be estimated from the quantity,
because the statistical noise of the quantity is very large.
Hence the fit of $\phi_\alpha( \vec{ r } )$ is carried out
by assuming the effective range
$ R = 7, 9 $ and $8$ for $ \alpha = 0, 1 $ and $ \sigma $ states.
The effective range depends on the state,
because $U_E(r,r^{'})$ in eq.~(1)
depends on the energy.
The fitting parameters are an overall constant of eq.~(2) and 
the relative momentum $p^2$.
Fig.~\ref{fig:fit} shows the fitting result $G_\alpha( r )$ for each state.
The fitting results for all states are consistent with the wave functions.
This figure illustrates 
that the wave function in $r>R$ for the $\sigma$ state can be
described by the general solution of the Helmholtz equation.
The consistency of the effective ranges is checked by 
the deviations of $\phi_\alpha ( r )$ from $G_\alpha ( r )$.
As expected, the deviation vanishes in $r > R$.

\begin{figure}[t!]
\centerline{\scalebox{.37}[.40]{\includegraphics{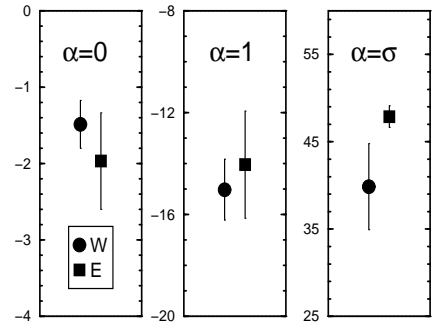}}}
\vspace{-10mm}
\caption{\label{fig:phct}
Scattering phase shift (degrees) obtained from wave function (W)
and energy (E) for ground, first excited and $\sigma$ states.
}
\vspace{-6mm}
\end{figure}

As shown in Fig.~\ref{fig:phct},
the phase shift is evaluated using
$p^2$ determined from the wave function.
In the figure the phase shift 
calculated with $p^2$ obtained from the energy 
$E_\alpha = 2\sqrt{ m^2 + p^2}$ is also presented.
All the phase shifts are consistent with the two methods.
The results obtained from the wave function for the ground and 
first excited states have smaller error than those obtained from 
the energies.
This feature is also seen in refs.~\cite{Balog} and \cite{WF}.
The wave function result for the $\sigma$ state, however,
has larger error.
The reason is not well understood at present.
In order to understand this feature, more detailed investigation is needed
for the wave function of the $\sigma$ state.

\section{Conclusions}
Calculation of the two-particle wave function
for the first excited and an unstable states 
is demonstrated with the diagonalization method.
It is found that the wave function in $r>R$ 
for an unstable state can be described
by the general solution of the Helmholtz equation as same as 
that for two-particle states,
and the phase shift is extracted from these wave functions.

\vspace{2mm}
The numerical calculations have been carried out 
on workstations at Center for Computational Sciences, University of Tsukuba.

%

%
%

\end{document}